\documentclass[10pt,journal,compsoc]{IEEEtran}

\usepackage{setspace} 
\usepackage[T1]{fontenc}
\usepackage{amsmath,amssymb,amsfonts,mathrsfs,bm}
\usepackage{amsthm}
\usepackage{cite}
\usepackage{array}
\usepackage[shortlabels]{enumitem}
\usepackage{graphicx}
\usepackage{url}
\usepackage{color}
\usepackage{algorithm,algorithmic}
\usepackage{multirow}
\usepackage{breqn}
\usepackage[table]{xcolor}
\usepackage[normalem]{ulem}
\usepackage{array}
\newcolumntype{L}[1]{>{\raggedright\let\newline\\\arraybackslash\hspace{0pt}}m{#1}}
\newcolumntype{C}[1]{>{\centering\let\newline\\\arraybackslash\hspace{0pt}}m{#1}}
\newcolumntype{R}[1]{>{\raggedleft\let\newline\\\arraybackslash\hspace{0pt}}m{#1}}
\usepackage{subfigure}

\usepackage{xparse}

\ifx\notloadhyperref\undefined
	\ifx\loadbibentry\undefined
		\usepackage[hidelinks]{hyperref} 
	\else
		\usepackage{bibentry}
		\makeatletter\let\saved@bibitem\@bibitem\makeatother
		\usepackage[hidelinks]{hyperref}
		\makeatletter\let\@bibitem\saved@bibitem\makeatother
	\fi
\else
\relax
\fi

\usepackage[capitalize]{cleveref}
\crefname{equation}{}{}
\Crefname{equation}{}{}
\crefname{claim}{claim}{claims}
\crefname{step}{step}{steps}
\crefname{line}{line}{lines}
\crefname{Theorem}{Theorem}{Theorems}
\crefname{Corollary}{Corollary}{Corollaries}
\crefname{Proposition}{Proposition}{Propositions}
\crefname{Lemma}{Lemma}{Lemmas}
\crefname{Definition}{Definition}{Definitions}
\crefname{Example}{Example}{Examples}
\crefname{Assumption}{Assumption}{Assumptions}
\crefname{Remark}{Remark}{Remarks}
\crefname{Theorem_A}{Theorem}{Theorems}
\crefname{Corollary_A}{Corollary}{Corollaries}
\crefname{Proposition_A}{Proposition}{Propositions}
\crefname{Lemma_A}{Lemma}{Lemmas}
\crefname{Definition_A}{Definition}{Definitions}

\ifx\useTheoremCounter\undefined
\newtheorem{Theorem}{Theorem}
\newtheorem{Corollary}{Corollary}
\newtheorem{Proposition}{Proposition}

\else

\fi

\theoremstyle{remark}

\newcommand{\Real}{\mathbb{R}}

\newcommand{\bB}{\mathbf{B}}
\newcommand{\bc}{\mathbf{c}}

\newcommand{\bF}{\mathbf{F}}

\newcommand{\bG}{\mathbf{G}}

\newcommand{\bP}{\mathbf{P}}

\newcommand{\bT}{\mathbf{T}}

\newcommand{\bbB}{\mathbb{B}}

\newcommand{\bbQ}{\mathbb{Q}}

\newcommand{\bbS}{\mathbb{S}}

\DeclareSymbolFont{bsfletters}{OT1}{cmss}{bx}{n}
\DeclareSymbolFont{ssfletters}{OT1}{cmss}{m}{n}
\DeclareMathSymbol{\bsfGamma}{0}{bsfletters}{'000}
\DeclareMathSymbol{\ssfGamma}{0}{ssfletters}{'000}
\DeclareMathSymbol{\bsfDelta}{0}{bsfletters}{'001}
\DeclareMathSymbol{\ssfDelta}{0}{ssfletters}{'001}
\DeclareMathSymbol{\bsfTheta}{0}{bsfletters}{'002}
\DeclareMathSymbol{\ssfTheta}{0}{ssfletters}{'002}
\DeclareMathSymbol{\bsfLambda}{0}{bsfletters}{'003}
\DeclareMathSymbol{\ssfLambda}{0}{ssfletters}{'003}
\DeclareMathSymbol{\bsfXi}{0}{bsfletters}{'004}
\DeclareMathSymbol{\ssfXi}{0}{ssfletters}{'004}
\DeclareMathSymbol{\bsfPi}{0}{bsfletters}{'005}
\DeclareMathSymbol{\ssfPi}{0}{ssfletters}{'005}
\DeclareMathSymbol{\bsfSigma}{0}{bsfletters}{'006}
\DeclareMathSymbol{\ssfSigma}{0}{ssfletters}{'006}
\DeclareMathSymbol{\bsfUpsilon}{0}{bsfletters}{'007}
\DeclareMathSymbol{\ssfUpsilon}{0}{ssfletters}{'007}
\DeclareMathSymbol{\bsfPhi}{0}{bsfletters}{'010}
\DeclareMathSymbol{\ssfPhi}{0}{ssfletters}{'010}
\DeclareMathSymbol{\bsfPsi}{0}{bsfletters}{'011}
\DeclareMathSymbol{\ssfPsi}{0}{ssfletters}{'011}
\DeclareMathSymbol{\bsfOmega}{0}{bsfletters}{'012}
\DeclareMathSymbol{\ssfOmega}{0}{ssfletters}{'012}

\DeclareMathOperator*{\argmax}{arg\,max}

\newcommand{\qednew}{\nobreak \ifvmode \relax \else
      \ifdim\lastskip<1.5em \hskip-\lastskip
      \hskip1.5em plus0em minus0.5em \fi \nobreak
      \vrule height0.75em width0.5em depth0.25em\fi}

\newcommand{\ofrac}[1]{{\frac{1}{#1}}}

\newcommand{\cond}[2]{\left. {#1}\, \middle| \, {#2} \right.}

\DeclareDocumentCommand \P { g d() g } {%
	\IfNoValueTF {#3} 
	{%
		\IfNoValueTF {#1} 
		{%
			\IfNoValueTF {#2}
			{%
				\mathbb{P}%
			}%
			{%
				\mathbb{P}\left({#2}\right)%
			}%
		}%
		{%
			\IfNoValueTF {#2}
			{%
				\mathbb{P}_{#1}%
			}%
			{%
				\mathbb{P}_{#1}\left({#2}\right)%
			}%
		}%
	}%
	{%
		\IfNoValueTF {#1} 
		{%
			\mathbb{P}\left(\cond{#2}{#3}\right)%
		}%
		{%
			\mathbb{P}_{#1}\left(\cond{#2}{#3}\right)%
		}%
	}%
}

\DeclareDocumentCommand \E { g o g } {%
	\IfNoValueTF {#3} 
	{%
		\IfNoValueTF {#1} 
		{%
			\IfNoValueTF {#2}
			{%
				\mathbb{E}%
			}%
			{%
				\mathbb{E}\left[{#2}\right]%
			}%
		}%
		{%
			\IfNoValueTF {#2}
			{%
				\mathbb{E}_{#1}%
			}%
			{%
				\mathbb{E}_{#1}\left[{#2}\right]%
			}%
		}%
	}%
	{%
		\IfNoValueTF {#1} 
		{%
			\mathbb{E}\left[\cond{#2}{#3}\right]%
		}%
		{%
			\mathbb{E}_{#1}\left[\cond{#2}{#3}\right]%
		}%
	}%
}

\definecolor{gray90}{gray}{0.9}

\newcommand{\msout}[1]{\text{\color{green} \sout{\ensuremath{#1}}}}
\newcommand{\del}[1]{{\color{green}\ifmmode \msout{#1}\else\sout{#1}\fi}}

\newcommand{\hide}[1]{}

\renewcommand{\figurename}{Fig.}
\graphicspath{{./Figures/}} 
\crefname{experiment}{experiment}{experiments}

\newcommand{\hH}{\widehat{H}}

\begin{document}

\title{Sequential Multi-Class Labeling in Crowdsourcing}
\author{Qiyu Kang, \IEEEmembership{Student Member,~IEEE,} and Wee Peng Tay, \IEEEmembership{Senior Member,~IEEE}
\thanks{This research is supported in part by the Singapore Ministry of Education Academic Research Fund Tier 2 grant MOE2014-T2-1-028.}
\thanks{The authors are with the School of Electrical and Electronic Engineering, Nanyang Technological University, Singapore. Email: KANG0080@e.ntu.edu.sg, wptay@ntu.edu.sg.} 
}

\IEEEtitleabstractindextext{
\begin{abstract}
We consider a crowdsourcing platform where workers' responses to questions posed by a crowdsourcer are used to determine the hidden state of a multi-class labeling problem. As workers may be unreliable, we propose to perform sequential questioning in which the questions posed to the workers are designed based on previous questions and answers. We propose a Partially-Observable Markov Decision Process (POMDP) framework to determine the best questioning strategy, subject to the crowdsourcer's budget constraint. As this POMDP formulation is in general intractable, we develop a suboptimal approach based on a $q$-ary Ulam-R\'enyi game. We also propose a sampling heuristic, which can be used in tandem with standard POMDP solvers, using our Ulam-R\'enyi strategy. We  demonstrate through simulations that our approaches outperform a non-sequential strategy based on error correction coding and which does not utilize workers' previous responses.  
\end{abstract}

}
\maketitle

\IEEEdisplaynontitleabstractindextext
\IEEEpeerreviewmaketitle

\IEEEraisesectionheading{\section{Introduction}\label{sec:intro}}

\IEEEPARstart{I}{n} a crowdsourcing platform, workers are given a task, like classifying or labeling an object in a picture, to perform. The crowdsourcer then makes a final decision based on the collective answers from all participating workers. For example, in\cite{bragg2013crowdsourcing}, crowdsourcing was used to produce taxonomies whose quality approaches that of human experts. Crowdsourcing platforms like Amazon Mechanical Turk \cite{snow2008cheap} typically has many participating workers. The goal is to make use of the abundance of workers to perform simple but tedious microtasks that do not require much domain expertise. However, workers may be highly unreliable \cite{Ross2010,Varshney2012}. Therefore, the microtasks are usually designed to be simple binary questions \cite{Vempaty2014a}. In this paper, we use the terms microtask and question interchangeably by assuming that workers are always required to answer a question posed by the crowdsourcer. 

To improve the reliability of the final decision, various inference algorithms and question allocation methods have been proposed. For example, in\cite{Karger2014,karger2011iterative}, the authors proposed a task assignment scheme using a bipartite graph to model the affinity of workers for different binary tasks and an iterative algorithm based on belief propagation to infer the final decision from the workers' responses.  In \cite{Khetan2016}, the authors proposed an iterative inference algorithm based on a spectral method. They assumed that tasks have different difficulties and workers have different reliabilities using a generalized Dawid-Skene model \cite{zhou2015regularized}. A multi-class labeling problem was considered in \cite{karger2013efficient}, in which an allocation algorithm is developed to assign tasks to different workers, and an inference method achieving an order-optimal redundancy-accuracy trade-off was developed. Redundancy here refers to assigning the same task to multiple workers and using a majority voting rule to infer the final answer. Some researchers have considered asking additional gold questions inserted amongst the actual work tasks to determine the reliability of workers \cite{oleson2011programmatic}. In \cite{singla2014near}, the authors presented a strategy to improve workers' reliability by teaching them classification rules. Several incentive strategies to motivate workers to participate in crowdsourcing have been studied in \cite{sun2014behavior,zhao2014crowdsource}. However, how to design the microtasks or questions for the workers has not addressed in the aforementioned works. 

In this paper, we assume that the crowdsourcer wishes to solve an $M$-ary multi-class labeling problem. From coding theory \cite{yao2007performance,wang2005distributed}, we know that it is possible to reduce the inference error through the use of error correction codes. The reference \cite{Vempaty2014a} developed an algorithm called Distributed Classification Fusion using Error-Correcting Codes (DCFECC) using error-correction coding to divide a single multi-class labeling task into binary questions, which are then assigned to the workers. The classification is then done by performing a Hamming distance decoding of all the workers' binary responses, which can be treated as the noisy versions of a coded message. We note that DCFECC is based on a non-feedback coding strategy. One can use a feedback coding strategy to further reduce the misclassification probability. In this paper, we consider the same multi-class labeling problem as \cite{Vempaty2014a} but now allow questions posed to the workers to depend on the workers' previous responses. Our goal is to develop a sequential approach to design questions for the crowdsourcing platform. We assume that at each questioning round, the crowdsourcer asks a group of workers a $q$-ary question (with $q\leq M$ and to be optimized), which we design based on the workers' answers in previous rounds. The $q$-ary question is further decomposed into binary questions for each worker using a similar error correction coding approach as DCFECC.

We formulate the sequential $q$-ary question design problem as a Partially-Observable Markov Decision Process (POMDP) \cite{ross2008online, pineau2003point, pineau2006anytime, silver2010monte,smallwood1973optimal}, which is however intractable due to its extremely large action space. This difficulty together with the well-known curse of dimensionality \cite{pineau2006anytime,silver2010monte} in POMDP makes optimally solving POMDP computationally intractable. We therefore develop approximate methods to solve the POMDP.

Our first strategy is to approximate the POMDP using a $q$-ary Ulam-R\'enyi game \cite{Cicalese2000a,Sarkissian1995,Pelc2002}. This is an iterative game in which one player chooses a state out of $M$ possible states, and another player asks the first player $q$-ary questions in order to determine the state chosen. The questions are asked sequentially, and a question at one iteration can be based on the responses to the questions from previous iterations. The first player may answer some of the questions posed wrongly. The first player's choice of the state corresponds to the true state of our multi-class labeling problem, while giving a wrong answer to a question corresponds to the unreliability of a worker in our crowdsourcing platform. We show how to find the best parameter $q$ in order to achieve an optimal trade-off between the classification accuracy and the number of iterations of questions required. As finding the optimal strategy for a Ulam-R\'enyi game is in general intractable, heuristics for solving a \emph{binary} Ulam-R\' enyi game have been proposed in \cite{Cicalese2000a,Sarkissian1995}. However, to the best of our knowledge, strategies for a general $q$-ary Ulam-R\' enyi game have been proposed only when the number of questions allowed is sufficiently large. In this paper, we propose an efficient heuristic, which is however suboptimal in general. 

Furthermore, based on our proposed Ulam-R\'enyi strategy, we further propose an action sampling strategy for the POMDP to sample a small subset of actions so that standard POMDP solvers like PBVI \cite{ pineau2003point, pineau2006anytime} and POMCP \cite{silver2010monte} can be used. Simulations suggest that this sampling strategy is better that uniform sampling.

A preliminary version of this paper appears in \cite{kang2017sequential}. In that conference paper, we introduced and presented a preliminary heuristic strategy for solving a $q$-ary Ulam-R\'enyi game, and applied that to a related crowdsourcing problem. In this paper, we have formulated our crowdsourcing problem as a POMDP, further refined our Ulam-R\'enyi strategy, and proposed a sampling strategy for use with POMDP solvers. More extensive simulation results are also included in this paper.

The rest of this paper is organized as follows. In \cref{sec:model}, we present our system model and assumptions, and formulate the crowdsourcing problem as a POMDP. In \cref{sec:Ulam}, we introduce the Ulam-R\'enyi game, and develop a novel heuristic to solve it. We then apply this strategy to an approximation of our POMDP formulation. In \cref{sec:POMDPSQS}, we describe our new action sampling strategy for traditional POMDP solvers. In \cref{sec:simulation}, we present simulations to compare the performance of our approaches with other strategies. \Cref{sec:conclusion} concludes the paper. 

\emph{Notations:} In this paper, we use $[a,b]$ to denote the set of integers $\{a,a+1,\ldots,b\}$, and $a_i^j$ to denote $(a_i, a_{i+1},\ldots,a_j)$. The notation $p(x\mid y)$ represents the conditional probability of the random variable $x$ given $y$, where $y$ can be a random variable or an event. We assume that all random variables and events are defined on a common probability space with probability measure $\P$ and expectation operator $\E$.

\section{Problem Formulation} \label{sec:model}

Consider a crowdsourcing platform that allows a crowdsourcer to pose questions or tasks to workers on the platform. In this paper, a crowdsourcer wishes to solve a $M$-ary multi-class labeling problem: determine a hidden state $H \in \bbS =[1,M]$ with the help of groups consisting of $N$ workers each. For example, the crowdsourcer may wish to classify an image of a dog into one of $M=9$ breeds: Dachshund, Irish Water Spaniel, Afghan Hound, Bull Terrier, Westie, Alaskan Malamute,  American Pit Bull Terrier, Fox Terrier and Rough Collie. In this case, we have $\bbS=\{1,2,...,9\}$, where each state corresponds to a particular dog breed in the order listed. However, since workers may not be experts in such image classification tasks, and can only answer simple multiple choice questions, the crowdsourcer asks each group of workers a sequence of $q$-ary questions, where $q \in [2, M]$.  A $q$-ary question can be represented as a $q$-tuple $\bT = (T_1,T_2,\ldots,T_{q})$ , where the sets $T_j$, $j\in [1,q]$, are pairwise disjoint subsets of $\bbS$. The $q$-tuple is interpreted as a question of the form ``Which one of the sets $T_1,\ldots,T_{q}$ does $H$ belong to?''. At each iteration, the $q$-ary question posed is designed based on the responses from previous iterations. Each $q$-ary question is assigned to the whole group of $N$ workers, and is further broken down into simpler microtasks that are assigned to each individual worker within the group. 

For a concrete illustration, consider again the example of classifying an image of a dog into one of nine breeds as described above. Suppose that the crowdsourcer uses tenary ($q=3$) questions. Then the question, ``Does the dog in the image have drop, erect or half-erect ears?'', can be represented as the tenary tuple $\bT = (\{1,2,3\},\{4,5,6\},\{7,8,9\})$ since $\{1,2,3\}$ corresponds to breeds with drop ears, $\{4,5,6\}$ corresponds to breeds with erect ears, and $\{7,8,9\}$ corresponds to breeds with half-erect ears. The crowdsourcer then assigns this question to the first group of workers. Suppose that he receives the answer ``The dog has drop ears'', then his next question could be ``Does the dog in the image have smooth, curly, or long and straight fur?'', which can be represented as $\bT = (\{1,4,7\},\{2,5,8\},\{3,6,9\})$. By iteratively asking different questions to groups of workers, the crowdsourcer is able to narrow down the exact dog breed in the image.

\subsection{Binary Microtasks with Code Matrix}\label{sec:sqsecc}

Suppose a $q$-ary question $\bT$ is posed to a worker group $g$, consisting of $N$ workers randomly chosen from the worker pool. Let $y_{g,k} \in [1,q]$ be the opinion or response of a worker $k \in[1,N]$ in the group if he is posed the question $\bT$. We assume that each worker $k$ has an associated reliability $\lambda_{g,k}$, drawn independently from a common distribution, with mean $\mu_q$. Let $\Lambda_t =(\lambda_{g,k})_{k=1}^N$. Conditioned on $H$ and $\Lambda_g$, we assume that workers' responses are independent from each other, and the response of worker $k$ in group $g$ has the following probability mass function:
\begin{align}
&p(y_{g,k} \mid H, \Lambda^{q}_{g}) =
		\left\{ 
		\begin{array}{ll}
				\lambda_{g,k} &\text{if $H \in T_{y_{g,k}}$},  \\
				\frac{1-\lambda_{g,k}}{q-1} &\text{otherwise}. 
		\end{array} 
		\right.  \label{eq:assumption1}
\end{align}

Note that we have assumed that the distribution generating $\lambda_{g,k}$ depends only on the number $q$ of possible responses, and not on the specific question $\bT$. This is because workers are typically non-experts without specific domain knowledge of the labeling problem assigned. Furthermore, we will typically assume in our simulations that $\mu_q$ is a decreasing function of $q$, i.e., workers become more unreliable as the question posed becomes more complex. Indeed, it has been observed that workers on crowdsourcing platforms typically do not have enough expertise to answer non-binary questions reliably \cite{branson2010visual}. Therefore, we first transform the $q$-ary question posed to each worker group into binary questions, which are then assigned to individual workers within the group. 

In the DCFECC approach proposed by \cite{Vempaty2014a}, a code matrix is designed to determine what questions are assigned to each worker to solve a $M$-ary labeling problem. For $N$ workers, a code matrix is a $M \times N$ binary matrix, where each column corresponds to a binary question for each worker, and each row corresponds to a code for each state of the $M$-ary labeling problem. The responses from all the $N$ workers can then be viewed as a noisy version of one of the rows in the code matrix, and a Hamming distance decoder is used to decode for the hidden state $H$.

In our proposed approach, a $q$-ary question $\bT=(T_1,\ldots,T_{q})$ is posed at each time to a worker group. We can interpret this as a $q$-ary labeling problem, and let $\bG_q$ be the corresponding $q \times N$ code matrix generated by DCFECC when applied to solving this $q$-ary labeling problem.
\begin{figure}[!htb]
\centering 
  \includegraphics[width=0.8\linewidth]{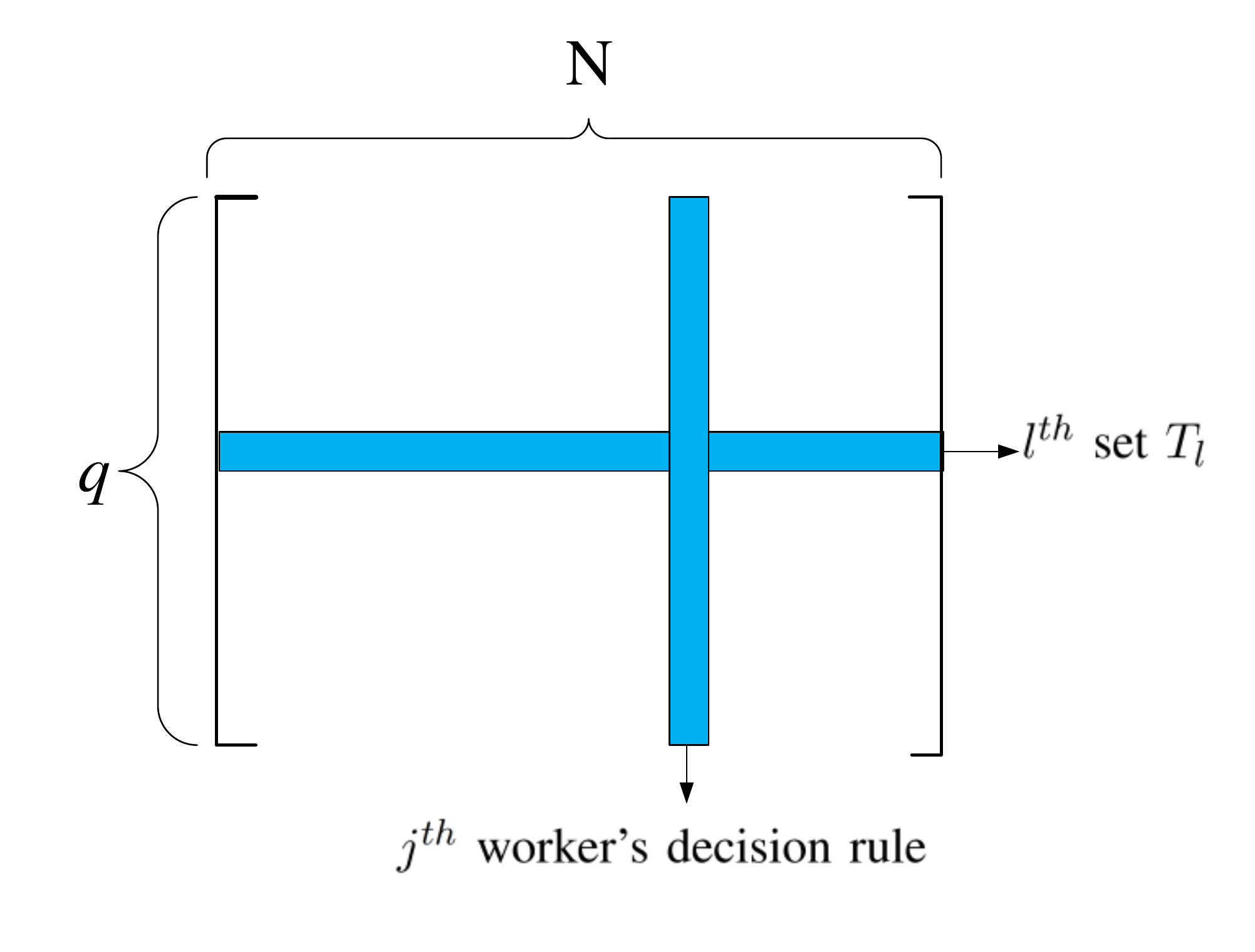}

 \caption{Code matrix representation of the DCFECC approach}
  \label{fig:matrix}
\end{figure}

The $j$-th column of the code matrix $\bG_q$ represents the binary question $\bB=(B_0,B_1)$ posed to the $j$-th worker, where $B_k=\{l\in [1,q]: \bG_q(l,j)=k\}$ for $k=0,1$ and $\bG_q(l,j)$ being the $(1,j)$-th element of $\bG_q$. Let $u \in \{0,1\}^N$ be the vector of responses from the $N$ workers. The $j$-th element in vector $u$ is set as bit $k$ if the $j$-th worker decides that $B_k$ contains the correct class. As workers may be unreliable, the responses from the $N$ workers can be viewed as a noisy version of one of the rows in $\bG_q$.
Using the same decision rule in \cite{yao2007performance}, we then decode $u$ to $T_l$ if the $l$-th row of $\bG_q$ has the smallest Hamming distance to $u$, with ties broken randomly. The set of responses whose Hamming distance to the $l$-th row of $\bG_q$ is minimal is called the decision region of $T_l$. 

Under the worker reliability model \cref{eq:assumption1}, it is shown in \cite{Vempaty2014a} that for any $\underline{i}=[i_1,i_2,\ldots,i_{N}] \in \{0,1\}^{N}$, we have
\begin{align} 
	&\P(u = \underline{i}){H\in T_l} \nonumber\\
	&= \prod^{N}_{j=1}\left[1-i_j+2(i_j-1)\left( \mu_q \bG_{l,j}+\frac{1-\mu_q}{q-1} \sum_{k\ne l} \bG_{k,j}  \right) \right],
\end{align} 
where $\bG_{l,j}$ is the $(l,j)$-th element of $\bG_q$. Let $o$ be the row number of $\bG_q$ found by the Hamming distance decoder. We have
\begin{align}
p(o\mid H\in T_l)=  \sum_{\underline{i}}\P(u = \underline{i}){H\in T_l}C^o(\underline{i}),
\end{align} 
where 
\begin{align*}
	C^{o}(\underline{i}) = 
	\left\{ 
		\begin{array}{ll}
		\ofrac{n_{\underline{i}}} \quad & \text{if } \underline{i} \text{ is in the decision region of } T_o, \\
		0\quad & \text{otherwise,} 
		\end{array} 
		\right.
\end{align*} 
with $n_{\underline{i}}$ being the number of decision regions $\underline{i}$ belongs to. We call the $q\times q$ matrix $\bP_q$ whose $(l,o)$-th element is $p(o\mid H\in T_l)$, the \emph{performance matrix} of the code matrix $\bG_q$. Similar to DCFECC, we choose the code matrix $\bG_q$ to minimize $\sum_{l=1}^q\sum_{o\ne l} p(o \mid H \in T_l)/q$.

\subsection{POMDP Formulation}

We assume that no prior information about $H$ is available. Therefore, we choose a uniform prior over $\bbS$ for $H$. At each time $t$, the crowdsourcer chooses an action $a_t$: to either pose a question to worker group $g$, or to stop and make a final decision. Let $\tau$ be the stopping time. Since most crowdsourcing platforms like Amazon Mechanical Turk require the crowdsourcer to financially compensate a worker for each task he completes, the crowdsourcer's expenditure increases with increasing $\tau$. Let $b$ be the budget of the crowdsourcer so that we have the constraint $\tau \leq b$. Let $\bbQ(q)$ be the set of all $q$-tuples $\bT$. For each time $t < \tau$, the crowdsourcer chooses a tuple $a_t\in \bbQ(q)$, poses it to the worker group $g$, and fuses all the answers from the workers in group $g$ to obtain an answer $o_t \in [1,q]$. The choice of the tuple $a_t$ is determined by a policy $\phi_t : \bbQ(q)^{t-1}\times [1,q]^{t-1} \mapsto \bbQ(q)$. At time $\tau$, the action or decision $a_\tau$ is given by a decision rule $\hH : \bbQ(q)^{\tau-1}\times [1,q]^{\tau-1} \mapsto \bbS$. Let $R = \P(\hH = H)$ be the probability that the decision $\hH$ is correct. We call $R$ the \emph{reliability} of the strategy. We define a \emph{policy} to be $(q, \tau, (\phi_t)_{t=1}^{\tau-1}, \hH)$.  Let the cost of each question posed be $\gamma$. The aim of the crowdsourcer is to 
\begin{align}\label{objective}
\begin{aligned}
\max \ & R-\gamma(\E\tau -1) \\
\text{subject to }& \tau \in [1,b],\\ 
\end{aligned}
\end{align}   
over all policies.

Let $p_0(h)=1/M$ for all $h\in \bbS$, and the belief at time $t$ be 
\begin{align*}
p_t(h) = \P(H=h){a_1^t, o_1^t},
\end{align*}
for each $h\in \bbS$. It can be shown that $R$ is a function of $p_\tau = (p_\tau(h))_{h\in \bbS}$ since the optimal decision rule at time $\tau$, for a given $(q, \tau, (\phi_t)_{t=1}^{\tau-1})$ is to choose $\hH = \argmax_{h\in \bbS} p_\tau(h)$. Thus, to solve \cref{objective}, it suffices to keep track of the sufficient statistics $p_t = (p_t(h))_{h\in \bbS}$, which can be computed sequentially via Bayes' rule as

\begin{align}
&p_t(h) =
		\left\{ 
		\begin{array}{ll}
				\frac{p(o_t\mid H=h,a_t=\bT)p_{t-1}(h)}{p(o_t \mid p_{t-1},a_t=\bT)}  &\text{if $h \in \cup^{q}_{j=1}T_j$},  \\
				p_{t-1}(h) &\text{otherwise}. 
		\end{array} 
		\right.  
\end{align}
where $p(o_t\mid H=h,a_t=\bT)$ is the $(l,o_t)$-th element of the performance matrix $\bP_q$ if $h\in T_l$, and 
\begin{align*}
&p(o_t\mid p_{t-1},a_t=\bT)\\ 
&= \frac{\sum_{h \in \cup^{q}_{j=1}T_j} p(o_t \mid H=h, a_t=\bT) p_{t-1}(h)}{\sum_{h \in \cup^{q}_{j=1}T_j}  p_{t-1}(h)}. 
\end{align*}

The formulation in \cref{objective} can be viewed as a POMDP in which we observe only the belief $p_t$ but not the hidden state $H$ at each time $t$. However, due to the large size of the action space $\bbQ(q)$, solving this POMDP exactly is intractable \cite{pineau2006anytime}. Furthermore, we need to optimize over $q\in [2,M]$. Most POMDP solvers, including PBVI\cite{pineau2003point, pineau2006anytime} and POMCP \cite{silver2010monte}, enumerate over all possible actions to find the optimal action. When the action space is large, such enumeration is no longer feasible. Several solvers \cite{seiler2015online,van2012efficient,kurniawati2012global} have even been proposed for continuous action spaces under some technical conditions. However, to the best of our knowledge, there is no general POMDP solver for large discrete action spaces. Therefore, in \cref{sec:Ulam}, we propose a heuristic approach based on the Ulam-R\'enyi game to solve \cref{objective}. In addition, as part of this heuristic approach, we obtain a sampling strategy to sample a small subset of actions from $\bbQ(q)$, which can be used in tandem with traditional POMDP solvers like PBVI and POMCP, in \cref{sec:POMDPSQS}.

\section{Ulam-R\'enyi Game Approach}
\label{sec:Ulam}

In this section, we first briefly introduce the concept of a Ulam-R\'enyi game, and propose a novel mixed-integer quadratic programming (MIQP) formulation to finding a heuristic strategy for the Ulam-R\'enyi game. We then approximate our POMDP in \cref{objective} with a Ulam-R\'enyi game and use our proposed Ulam-R\'enyi strategy to solve it. Finally, we propose an action sampling approach based on the Ulam-R\'enyi game, which can be used in tandem with standard POMDP solvers.

\subsection{A Heuristic Strategy for the \texorpdfstring{$q$}{q}-ary Ulam-R\'enyi Game}

In a $(M,q,e)$ Ulam-R\'enyi game \cite{Cicalese2000a,Sarkissian1995,Pelc2002}, a responder first chooses a number $H \in \bbS$, and a questioner asks the responder a sequence of $q$-ary questions to determine $H$. The responder is allowed to lie at most $e$ times. The questioner chooses his question at each step based on the previous responses, and his goal is to determine $H$ with the minimum number of questions. The Ulam-R\'enyi game is similar to our crowdsourcing problem: the questioner is our crowdsourcer, while the responder is a group of workers. The difference is that in our crowdsourcing formulation, the workers do not know $H$ a priori, therefore a wrong answer is modeled to be stochastic and there is no upper limit $e$ to the number of wrong answers the workers can give. In the following, we first give a brief overview of the Ulam-R\'enyi game and present a heuristic strategy to solve it. We then apply this strategy to our problem.

In a Ulam-R\'enyi game, we define a \emph{game status} $\sigma = (A_0,A_1,\ldots,A_e)$, where $A_i$, $i\in[0,e]$, are pairwise disjoint subsets of $\bbS$. The set $A_i$ contains all states $m\in \bbS$ that are potentially $H$ if the responder has lied exactly $i$ times. The initial game status $\sigma$ is given by $(\bbS,\emptyset,\ldots,\emptyset)$. At each question, if the current status is $\sigma = (A_0,A_1,\ldots,A_e)$ and the answer to the question $\bT = (T_1,T_1,\ldots,T_{q})$ is $j$, where $T_1,\ldots,T_{q}$ are pairwise disjoint sets and $\cup^{q}_{j=1}T_j = \cup^{e}_{i=0}A_i$, then the game status is updated as:
\begin{align}
		\sigma^j = ( & A_0\cap T_j, (A_0 \backslash T_j)\cup(A_1\cap T_j),\nonumber \\
		&\ldots,(A_{e-1}\backslash T_j)\cup(A_e\cap T_j)).\label{sigmaj}
\end{align}

For example, in the $(4,2,1)$ Ulam-R\'enyi game, the initial status is $\sigma = (\{1,2,3,4\},\emptyset)$. Suppose the responder chooses $H = 1$ and the questioner's first question is $\bT=(T_1,T_2)$ where $T_1=\{1,2\}$ and $T_2=\{3,4\}$. The responder can lie to this question and answer ``$H$ is in $T_2$'', and the questioner updates the game status $\sigma$ to $(\{3,4\},\{1,2\})$. This game status tells the questioner that if $H\in\{1,2\}$, then the responder has lied once, whereas if $H\in\{3,4\}$, then the responder has not lied. Suppose the second question posed to the responder is $\bT=(\{1,3\},\{2,4\})$. Since the responder has already lied once, he can only give the answer ``$H$ is in $T_1$''. The game status then becomes $\sigma = (\{3\},\{1,4\})$. Finally, if the questioner poses the question $\bT=(\{1\},\{3,4\})$ followed by another question $\bT=(\{1\},\{3\})$, the game status $\sigma = ( \emptyset, \{1\})$ is reached, at which point the questioner determines that $H=1$.

In general, it is intractable to find an optimal strategy for a $(M,q,e)$ Ulam-R\'enyi game \cite{Muthukrishnan1994,Cicalese2000a}. The reference \cite{Cicalese2007} proposed a search strategy that is near optimal when the number of questions is sufficiently large. In the following, we propose a heuristic adaptive strategy. For each $i\in[0,e]$, let $|A_i|$ be the number of states in $A_i$. Let $|\sigma|=(|A_0|,|A_1|,\ldots,|A_e|)$, which is called the type of $\sigma$\cite{Cicalese2007}. Same as \cite{Muthukrishnan1994}, we define the weight of an element $x\in A_i$ in status $\sigma$ with $w$ iterations left as 
\begin{align*}
W_{w}(i)=\sum_{k=0}^{e-i} \binom{w}{k}(q-1)^k,
\end{align*}
and the weight of the status $\sigma$ as the cumulative weight of all the elements in it, i.e., 
\begin{align*}
V_{w}(|\sigma|)= \sum_{i=0}^{e} |A_i| W_w(i).
\end{align*}
The weight $W_w(i)$ of an element $x\in A_i$ has an intuitive interpretation: if $x=H$, then there are $\binom{w}{k}(q-1)^k$ ways for the responder to lie $k$ times in the remaining $w$ iterations, so $W_w(i)$ is the total number of ways of lying for the responder. A game status $\sigma$ with high $V_w(|\sigma|)$ implies that the responder has more ways to obfuscate the true state $H$, making it more difficult for the questioner to determine $H$.

From Proposition 3.1 of \cite{Cicalese2013}, we have $V_{w}(|\sigma|) = \sum_{j=1}^{q} V_{w-1}(|\sigma^j|)$. Furthermore, since the responder lies at most $e$ times, Proposition 3.1 of \cite{Cicalese2013} and Theorem 3.1 of \cite{Muthukrishnan1994} show that the questioner is guaranteed to find $H$ in $w$ questions only if $V_w(|\sigma|)\leq q^w$. Consider the case where $V_w(|\sigma|)=q^w$ for some $w$. If the next question yields $V_{w-1}(|\sigma^j|) > q^{w-1}$ for some $j \in [1,q]$, the responder can answer $j$ to prevent the questioner from finding $H$. Therefore, the questioner should use a question $\bT$ such that $V_{w-1}(|\sigma^j|)$ is approximately the same for all $j\in[1,q]$. For each $j\in[1,q]$, we let
\begin{align*}
 	T_j &= \bigcup_{i=0}^e T_{j,i},
\end{align*}
where $T_{j,i} = T_j\cap A_i$, $\cup_{j=1}^{q} T_{j,i} = A_{i}$ and $T_{j,i}$, $j\in[1,q]$, are pairwise disjoint sets. We use $|T_j|$ to denote $(|T_{j,0}|,|T_{j,1}|,...,|T_{j,e}|)$. 

We observe from our experiments that $W_w(i)$ is usually much larger for smaller values of $i$. For example,  when $q=4$ and $e=7$, we have $W_{10}(0)=497,452$ while $W_{10}(7)=1$. This suggests the following heuristic: We first determine the value of $|T_{j,i}|$ for small values of $i$ and then use $|T_{j,i}|$ with larger values of $i$ to even out the weights $V_{w-1}(|\sigma^j|)$ for $j\in[1,q]$. 

Suppose $|T_{j,0}|,\ldots\,|T_{j,i-1}|$, for all $j\in [1,q]$ have been determined. Subsequently for $i\in[0,e]$, let $|\sigma_{i}^j|=(|A_{0}^j|,\ldots,|A_{i}^j|,|A_{i+1}^j|,0,\ldots,0)$, where 
\begin{align*}
|A_i^j| = |A_{i-1}|-|T_{j,i-1}|+|T_{j,i}|,
\end{align*}
with $|A_{-1}|=|T_{j,-1}|=|T_{j,i+1}|=0$. We propose to solve the following optimization problem:
\begin{align} \label{obj1}
\begin{aligned}
    \min_{(T_{j,i})_{j=1}^{q}}\ &\sum_{j,l} (V_{w-1}(|\sigma_i^j|) - V_{w-1}(|\sigma_i^l|))^2 \\
	\text{subject to }&|T_{j,i}|\in [0,|A_i|],\ \sum_{i=1}^{q}|T_{j,i}|=|A_i|.
\end{aligned}	
\end{align}
Let $V_j=\sum_{k=0}^{i-1}W_{w-1}(k)|A_i^j|-W_{w-1}(i)|T_{j,i-1}|$. The objective function in \cref{obj1} can be written as
\begin{align*}
\sum_{j,l}\left( V_j-V_l + (W_{w-1}(i)-W_{w-1}(i+1)) (|T_{j,i}|-|T_{l,i}|)\right)^2,
\end{align*}
where $W_{w-1}(e+1) = 0$. Then, the optimization problem \cref{obj1} can be reformulated as a MIQP as follows:
\begin{align}
\begin{aligned} \label{obj4_1}
			\min_{x} &\ofrac{2}x^T\bF x +\bc^Tx \\
			\text{subject to } &\mathbf{1}^{T}x= |A_i|\\ 
								&x_j \in [0,|A_i|], j \in [1,q]
		\end{aligned}
		\end{align}			
where $\bF \in \Real^{q\times q}$ has diagonal entries  $(q-1)\times \big(W_{w-1}(i)-W_{w-1}(i+1)\big)$, and for $j\ne k$, its $(j,k)$-th entry is $W_{w-1}(i+1)-W_{w-1}(i)$. The vector $\bc\in R^q$ has $j$-th entry $c_j=\sum_{k=0}^{q-1}(V_j - V_k)$, and $x_j$ denotes the $j$-th entry of vector $x$.

After obtaining the optimal $(|T_{j,i}|)_{j\in [1,q]}$ from solving \cref{obj4_1}, $A_i$ is arbitrarily partitioned by randomly choosing $|T_{j,i}|$ elements from it to form $T_{j,i}$. The above heuristic is presented in detail in Algorithm \ref{algor:heuris}.

\begin{algorithm}[!tb]
\caption{Heuristic strategy for a $(M,q,e)$ Ulam-R\'enyi game.} 
\label{algor:heuris} 
\begin{algorithmic}
\REQUIRE Current status $\sigma=(A_0,A_1,A_2,\ldots,A_e)$, with $w$ questions left.
\ENSURE  $\textbf{T}$
\STATE Let $m$ denote the smallest index s.t $|A_m| \ne 0 $.
 Let $\delta = \bmod (|A_m|,q)$, and $\omega = \lfloor \frac{|A_m|}{q} \rfloor$.
\FOR{$j=1$ to $\delta$}
\STATE Set $|T_{j,m}| = \omega+1$.
\ENDFOR 
\FOR{$j=\delta+1$ to $q$} 
\STATE Set $|T_{j,m}| = \omega$.
\ENDFOR 
\STATE Set $|T_{j,i}|=0$ for all $j\in [1,q]$ and $i \in [0,m-1]$.
\FOR{$i=m+1$ to $e$} 
	\IF{$|A_i| \ne 0$} 
		\STATE Solve \cref{obj4_1} to obtain an optimal $x$.
		\FOR{$j=1$ to $q$}
			\STATE Set $|T_{j,i}| = x_j$.
		\ENDFOR
	\ELSE
		\STATE Set $|T_{j,i}|=0$, for all $j\in [1,q]$.
	\ENDIF
\ENDFOR
\STATE For all $i\in [0,e]$, $A_i$ is arbitrarily partitioned by randomly choosing $|T_{j,i}|$ elements from it to form $T_{j,i}$, for all $j\in [1,q]$.
\end{algorithmic} 
\end{algorithm}

Let $B(q,e)$ be the minimum number of questions required to determine $H$ if at most $e$ of the questions have wrong responses. We propose a heuristic to compute $B(q,e)$ in Algorithm \ref{algor:findB}. The Ulam-R\'enyi game ends when the game status contains only a single state. Given the initial game status $\sigma = \{S,\emptyset,\ldots,\emptyset\}$ and the number of questions $w$, we use Algorithm \ref{algor:heuris} to generate a $q$-ary tree whose depth is $w$ and root corresponds to the initial game status. Each node in the tree corresponds to a game status. The $j$-th child of a node $\sigma$ is the status $\sigma^{j}$ updated from answer $j$ as defined in \eqref{sigmaj}. To find $B(q,e)$ then corresponds to finding the minimum $w$ that leads to a $q$-ary tree in which every leaf node has a game status that contains only one state. We initialize $w$ as $N_{\min}(e)=\min\{n \mid M\sum_{j=0}^{e}\binom{n}{j} (q-1)^{j} \le q^n\}$ since $B(q,e) \geq N_{\min}(e)$\cite{Muthukrishnan1994,Cicalese2007}. We then increment $w$ by one each time until the game status at every leaf node contains only one state. We call this a $(M,q,e)$ \emph{Ulam-R\'enyi tree}. The above procedure is summarized in Algorithm \ref{algor:findB}.

\begin{algorithm}[!tb]
\caption{Heuristic to compute $B(q,e)$.}
\label{algor:findB} 
\begin{algorithmic}
\REQUIRE $M,q,e$ 
\ENSURE  $B(q,e)$
\STATE Set $w = N_{\min} (e)=\min\{n \mid M\sum_{j=0}^{e}\binom{n}{j} (q-1)^{j} \le q^n  \}$.
\WHILE{1}
\STATE Use Algorithm \ref{algor:heuris} to generate a $q$-ary Ulam-R\'enyi tree whose depth is $w$ and root is the initial status.
	\IF{every leaf node's game status contains only one state}
	\STATE	break;
	\ELSE
	\STATE
		$w = w+1$;
	\ENDIF
\ENDWHILE
\RETURN $B(q,e) = w;$
\end{algorithmic} 
\end{algorithm}

To verify the performance of our heuristics, we compare our estimated $B(2,e)$ with that of the optimal $(M, 2, e)$ Ulam-R\'enyi game strategy in \cite{Cicalese2000a}, where $M=2^m$ is a power of 2. (Note that finding optimal strategies for $q>2$ is an open problem.)  The results are shown in Table \ref{tab:table1}, in which the numbers in brackets indicate the values found by \cite{Cicalese2000a} if our method differs from the optimal values. We see that our algorithm computes $B(2,e)$ correctly in most cases.

\begin{table}[!htb]
\caption{Estimated $B(2,e)$ of the $(M, 2, e)$ Ulam-R\'enyi game where $M=2^m$.}
\label{tab:table1} 
\centering
\begin{tabular}{c| c c c c c c c c}
 \hline
 \multirow{2}{*}{$m$} &\multicolumn{8}{c}{$e$} \\
 \cline{2-9}
  & 1 &2 & 3 &4 &5 &6 &7 &8\\
\hline

	1 & 3 & 5 & 7 & 9 & 11 & 13 & 15 & 17 \\ 
	2 & 5 & 8 & 11 & 14 & 17 & 20 & 23 & 26 \\ 
	3 & 6 & 9 & 12 & 15 & 18 & 21 & 24 & 27 \\ 
	4 & 7 & 10 & 13 & 16 & 19 & 22 & 25 & 28 \\ 
	5 & 9 & 12 & 15 & 18 & 21 & 24 & 27 & 30 \\ 
	6 & 10 & 13 & 16 & 19 & 22 & 25 & 28 & 31 \\ 
	7 & 11 & 14 & 17 & 20 & 23 & 26 & 29 & 32 \\
	8 & 12 & 15 & 18 & 21 & 24 & 27 & 30 & 33 \\
	9 & 13 & 17 & 20 & 23 & 26 & 29 & 32 & 35 \\
	10 & 14 & 18 & 21 & 24 & 27 & 30 & 33 & 36 \\
	11 & 15 & 19 & 22 & 25 & 28 & 31 & 34 & 37 \\
	12 & 17 & 20 & 23 & 27 & 30 & 33 & 36 & 39 \\
	13 & 18 & 21 & 25 & 28 & 31 & 34 & 37 & 40 \\
	14 & 19 & 22 & 26 & 29 & 32 & 35 & 39 (38) & 42 (41) \\
	15 & 20 & 24 & 27 & 30 & 34 & 37 & 40 & 43 \\
	16 & 21 & 25 & 28 & 32 & 35 & 38 & 41 & 44 \\ \hline
\end{tabular}
\end{table}

\subsection{POMDP Approximation}\label{ssec:approximateulamrenyi}

We now approximate the POMDP in \cref{objective} using a $(M, q, e)$ Ulam-R\'enyi game, and optimize over $(q,e)$. For a given $(q,e)$, we treat the crowdsourcer as the questioner in the $(M,q,e)$ Ulam-R\'enyi game, and each group of workers as the responder. We can think of the group of workers giving an incorrect response after decoding via the code matrix as the responder in the Ulam-R\'enyi game lying. However, in our crowdsourcing problem, the worker groups are not constrained to making at most $e$ errors. Therefore, after $B(q,e)$ questions, the crowdsourcer is not guaranteed to infer the correct hidden state $H$. Let $R(q,e)$ be the probability that the crowdsourcer finds the correct $H$ by applying our proposed heuristic in \cref{algor:heuris} iteratively till the game status contains only a single state, which is then declared to be the inferred hidden state $\hH$. Let $\hat{B}(q,e)$ be the maximum number of iterations required as computed in \cref{algor:findB}. Then we have $\tau \leq \hat{B}(q,e)+1$. 

Recall that $\bP_q$ is the performance matrix of the DCFECC code matrix $\bG_q$ used to transform the $q$-ary question $\bT$ into binary microtasks for each worker. Let $\underline{p} = \min_{i\in [1,q]} \bP_{q_{i,i}}$, where $\bP_{q_{i,i}}$ is the $(i,i)$-th element of $\bP_q$ and denotes the least probability that the question is answered correctly. Since $R(q,e)$ is the probability that all the worker groups make at most $e$ errors in at most $\hat{B}(q,e)$ questions, we have
\begin{align} 
	R(q,e) \geq \hat{R}(q,e) = \sum_{k=0}^{e} \binom{\hat{B}(q,e)}{k} (1-\underline{p})^k \underline{p}^{\hat{B}(q,e)-k},\label{lowerbound:Rqe}
\end{align}%
where the inequality follows from the fact that $\sum_{k=0}^{e} \binom{x}{l} (1-\underline{p})^k \underline{p}^{(x-k)}$ is a non-increasing function of $x$.

In place of the problem \cref{objective}, we consider instead the following optimization problem: 
\begin{align}\label{ULObj}
	\begin{aligned}
	\max_{q,e}& \ L(q,e) = \hat{R}(q,e)-\gamma \hat{B}(q,e) \\
	\text{subject to } & \hat{B}(q,e)\leq b-1,\\ 
	& q\in [2,M],\ e \in [0,\infty).
	\end{aligned}
\end{align}
Since the problem \cref{ULObj} is a discrete optimization problem, it is in general computationally intractable to find the exact optimum. For $M$ that is not too large, the solution space is relatively small and the solution can be found through enumeration. For large $M$, an integer programming method like branch and bound \cite{land1960automatic, lawler1966branch} or local search strategy \cite{ahuja2002survey} can be used to find an approximate solution. 

Let $(q^*,e^*)$ be the optimal solution of \cref{ULObj}. We call the $(M,q^*,e^*)$ Ulam-R\'enyi game strategy found using \cref{algor:heuris} together with the use of the DCFECC code matrix to transform the $q^*$-ary questions into binary questions, the Ulam-R\'enyi sequential questioning strategy (URSQS). 

\section{Sampling Using Ulam-R\'enyi Tree}\label{sec:POMDPSQS}

To solve the POMDP \cref{objective} directly using standard POMDP solvers is computationally expensive without using some sampling strategy to reduce the size of the action space. A typical sampling strategy is uniform random sampling. In this section, we propose a new sampling strategy based on the Ulam-R\'enyi tree.

Let $\hat{\bbQ}$ denote the sampled subset of the action space, where $|\hat{\bbQ}|$ is predetermined. Let $(q^*,e^*)$ be the solution of \cref{ULObj}. To sample each action in $\hat{\bbQ}$, use \cref{algor:findB} to find a $(M,q^*,e^*)$ Ulam-R\'enyi tree. We then randomly sample $|\hat{\bbQ}|$ nodes in the tree, and include the corresponding questions that generated these nodes into $\hat{\bbQ}$. We call this the Ulam-R\'enyi tree (URT) sampling strategy.

\section{Simulations and Performance Evaluation}\label{sec:simulation}

In this section, we perform simulations and compare the performance of our proposed approaches with the non-sequential DCFECC approach, and strategies found using the standard POMDP solvers PBVI \cite{pineau2003point, pineau2006anytime} and POMCP \cite{silver2010monte} with uniform sampling.

We perform simulations using $N = 10$, and  $\mu_q = rq^{-0.2}$. We vary the number of classes $M$, the budget $b$, cost of each question $\gamma$, and the parameter $r$, which controls the workers' reliability via $\mu_q$. 
We choose $\mu_q = rq^{-0.2}$ to be a decreasing function of $q$, as a worker's reliability typically decreases with increasing $q$ \cite{branson2010visual}. The parameter $r$ controls the workers' general reliability. A larger $r$ indicates that the workers are more reliable.
In each simulation, we perform 50,000 trials using URSQS, DCFECC, and the PBVI solver. Since DCFECC is a non-sequential strategy, we assume that it uses $N(b-1)$ workers, which is the maximum total number of workers used by URSQS and the POMDP solvers. We perform $2000$ trials when using the POMCP solver since it is significantly more computationally complex. To compare the performances of different approaches, we compute the empirical average, over all the trials, of the objective function in \cref{objective}. For convenience, this average is referred to as the \emph{average reward}.


\subsection{Comparison of Sampling Strategies}

When using the standard POMDP solvers, we choose the action space sample size $|\hat{\bbQ}|$ according to the number of classes $M$. For the PBVI solver, we also randomly sample a finite set of beliefs, $\bbB$. 

The performance of PBVI depends on the size of $\bbB$ relative to $M$. PBVI first computes the solutions for all samples in $\bbB$ to approximate the solutions for the full belief space. With denser belief samples, PBVI can converge to the optimal value \cite{pineau2003point}. However, since the full belief space grows exponentially with $M$, it becomes computationally intractable for large $M$. Due to computation limits, we are forced to reduce $|\bbB|/M$ for large $M$, leading to worst performance compared to POMCP, which only considers the current belief encountered in the process \cite{ross2008online, silver2010monte}. 

The full action space $\bbQ$ also grows exponentially with $M$. From \cref{tab:tableQ32}, we observe that the performance of both PBVI and POMCP deteriorate when $|\hat{\bbQ}|$ is too small. In our simulations, keeping within our computational limits, we set the values of $|\hat{\bbQ}|$ and $|\bbB|$ for different $M$ as shown in \cref{tab:tableaction}.

\begin{table}[!htb]
 \caption{Average reward for different $|\hat{\bbQ}|$ with $M = 32$, $r = 0.75$, $b = 9$,$\gamma = 0.05$ and URT sampling.}
\label{tab:tableQ32} 
\centering
\begin{tabular}{| c| c | c | c| c |}
 \hline 
	$|\hat{\bbQ}|$ & 2  & 50 & 100 & 300\\  \hline
	PBVI &0.122  &0.433 & 0.485 & 0.494 \\  \hline
	POMCP &0.102  &0.453 & 0.488 & 0.502 \\  \hline
\end{tabular}
\end{table}

\begin{table}[!htb]
 \caption{$|\hat{\bbQ}|$  and $|\bbB|$ for different $M$.}
\label{tab:tableaction} 
\centering
\begin{tabular}{| c| c | c | c | c | c |}
 \hline 
	$M$ & 8 & 16 & 32 & 64 & 128\\  \hline
	$|\hat{\bbQ}|$ & 100 & 200 & 300 & 300 & 300 \\  \hline
	$|\bbB|$ & 2000 & 4000 & 6000 & 7000 & 8000 \\ \hline
\end{tabular}
\end{table}

We now compare the performance of different sampling strategies when used in tandem with PBVI and POMCP. We vary $M$ while setting $r = 0.75$, $b = 9$ and $\gamma = 0.05$. 

\begin{figure}[!htb]
\centering 
  \includegraphics[width=1\linewidth]{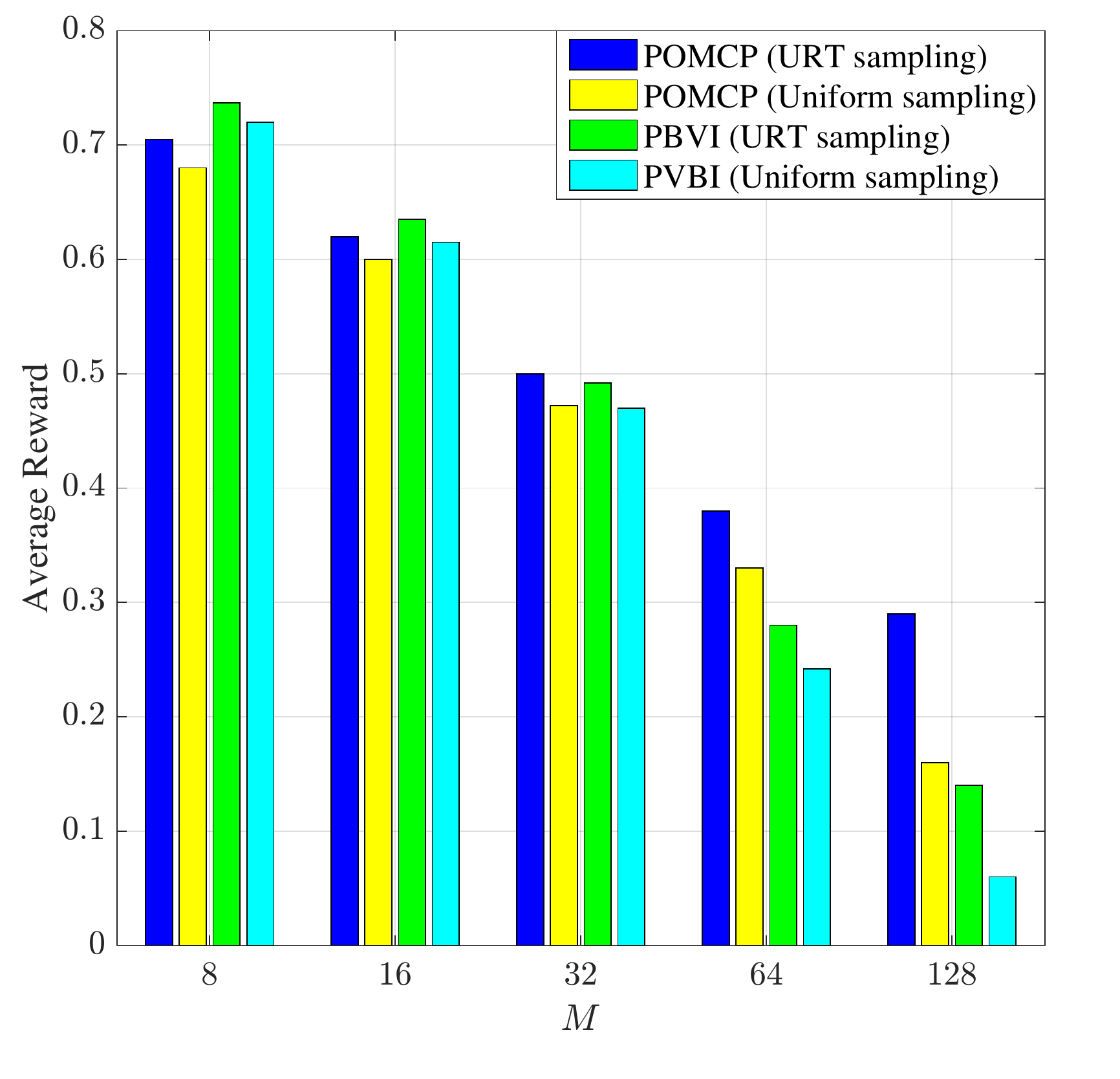}
  \vspace*{-8mm}
 \caption{Performance comparison between uniform sampling and URT sampling with $r = 0.75$, $b = 9$, $\gamma = 0.05$ and different $M$.}
  \label{fig:actionsampling}
\end{figure}

From \cref{fig:actionsampling}, we observe that the URT sampling produces a higher reward compared to uniform sampling for both PBVI and POMCP. In particular, URT sampling significantly outperforms uniform sampling when $M$ is large.
Therefore, in the remaining simulations, we adopt the URT sampling strategy when using PBVI and POMCP.

\subsection{Varying \texorpdfstring{$M$}{M}}\label{subsec:VaryingM}

\begin{figure}[!htb]
\centering
  \includegraphics[width=1\linewidth]{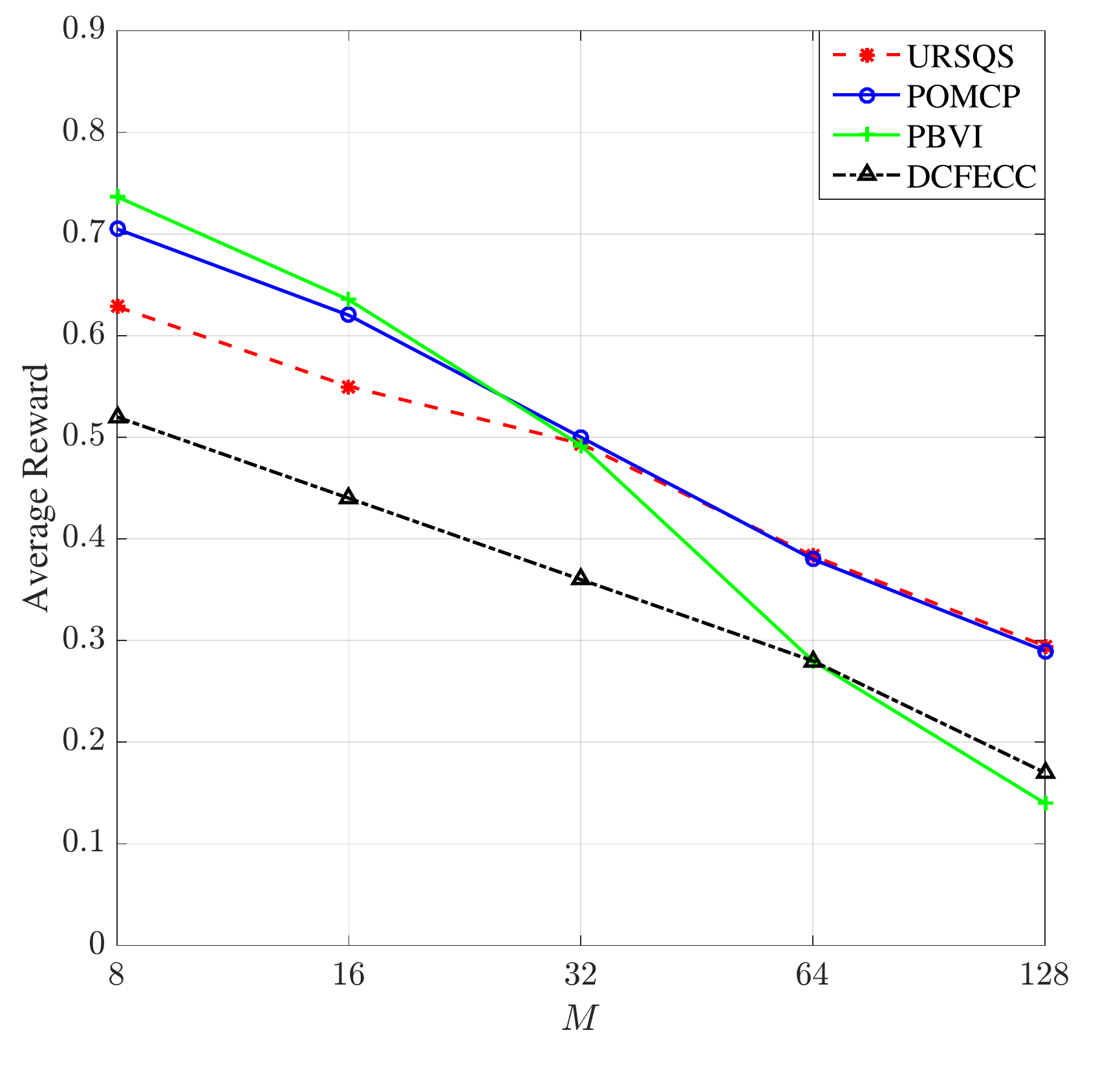}
  \centering
  \vspace*{-8mm}
 \caption{Performance comparison between strategies with $r=0.75$, $b = 9$, $\gamma=0.05$ and different $M$.}
  \label{fig:M}
\end{figure}

We next fix $r = 0.75$, $\gamma =0.05$, $b = 9$, and vary the number of classes $M$. The results are shown in \cref{fig:M}. We observe the following.
\begin{itemize}
	\item When $M$ is small, PBVI and POMCP with URT sampling achieve better rewards than URSQS. When $M$ is large, the rewards for POMCP and URSQS are comparable, while PBVI performs worse. 
	\item DCFECC is a non-sequential strategy and has the worst reward in most cases, since it does not make use of the answers from earlier workers to help design questions for the later workers. 
	\item We note that PBVI performs the worst when $M=128$ is large. One reason for this phenomenon is that PBVI is an offline approximate solver for POMDP, suffering from the curse of dimensionality as we have mentioned above. Also, as the full action space $\bbQ$ grows exponentially with $M$, sampling a small subset $\hat{\bbQ}$ cannot guarantee a good optimal solution. We observe that POMCP is better than PBVI when $M$ is large. 
\end{itemize}

To compare the computation complexity of the different methods under different $M$ values, we perform experiments on a setup with an Intel Xeon CPU E3-1226 v3 and 16GB RAM. The computation time is divided into two parts:
\begin{enumerate}[(i)]
	\item Offline computation: 	In all the methods, we first solve \cref{ULObj} to find the optimal $(q^*,e^*)$ and then use \cref{algor:findB} to construct a Ulam-R\'enyi tree. POMCP then performs URT sampling and stores the sample points. PBVI performs URT sampling and computes the solutions for all the samples, and is thus the most computationally expensive. URSQS does not require any additional offline processing, and thus have the lowest offline computation overhead. The computation times are shown in \cref{tab:tabletime2}. 
	\item Online computation: The average running time for each method over multiple trials and for different $M$ values are shown in \cref{tab:tabletime1}. We note that the computation complexity for POMCP is significantly higher than the other methods, although it produces the best average reward. PBVI has the lowest running time but its performance deteriorates for large $M$. URSQS has comparable performance as POMCP but much lower computation time. Our experiments suggest that PBVI with URT sampling is the recommended approach for small $M$ while URSQS is the recommended approach for large $M$.
\end{enumerate}	

\begin{table}[!htb]
 \caption{Average offline computation times for different strategies with $r=0.75$, $b = 9$, $\gamma=0.05$ and different $M$.} 
\label{tab:tabletime2} 
\centering
\begin{tabular}{| c| c | c | c | c | c |}
 \hline 
	          & $8$   &  $16$   & $32$      & $64$ & $128$\\  \hline
	URSQS (s)  & $386$ & $651$  & $1295$ &  $1492$ & $2405$ \\  \hline
	PBVI (s)   & $518$ & $1244$  & $2645$ &  $3602$ & $7065$ \\ \hline
	POMCP (s)  & $387$ & $653$  & $1300$ &  $1497$ & $2411$ \\ \hline
\end{tabular}
\end{table}

\begin{table}[!htb]
 \caption{Average online computation times for different strategies with $r=0.75$, $b = 9$, $\gamma=0.05$ and different $M$.} 
\label{tab:tabletime1} 
\centering
\begin{tabular}{| c| c | c | c | c | c |}
 \hline 
	          & $8$    &  $16$   & $32$      & $64$ & $128$\\  \hline
	URSQS (ms)& $7.04$ & $8.00$  & $16.34$ &  $18.15$ & $19.31$ \\  \hline
	PBVI (ms) & $1.81$ & $2.34$  & $2.78$ &  $3.95$ & $5.30$ \\ \hline
	POMCP (s) & $30$   & $124$  & $259$ &  $295$ & $352$ \\ \hline
\end{tabular}
\end{table}

\subsection{Varying \texorpdfstring{$b$}{b}}

\begin{figure}[!htb]
 \includegraphics[width=1\linewidth]{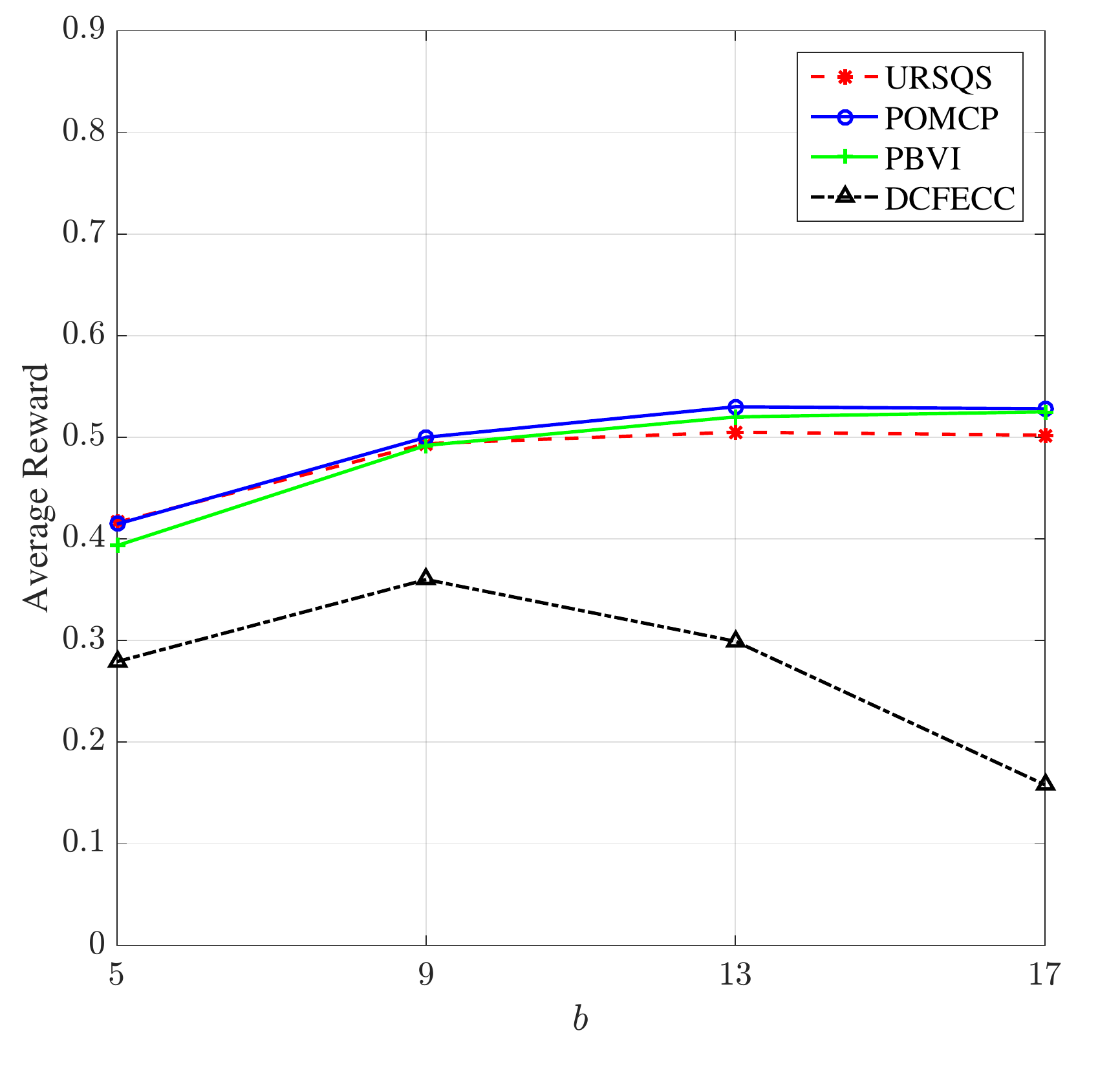}
\vspace*{-8mm}
 \caption{Performance comparison between strategies with $M = 32$, $r=0.75$, $\gamma=0.05$ and different $b$.}
 \label{fig:Budget}
\end{figure}

We next fix $M = 32$, $r = 0.75$, $\gamma = 0.5$ and vary $b$, the budget of the crowdsourcer. The results are shown in \cref{fig:Budget}. We observe the following:
 \begin{itemize}
	\item When the budget $b$ is relatively small, the rewards for POMCP, PBVI and URSQS increase with increasing budgets, as \cref{fig:Budget} depicts. This is because with larger budgets, more questions can be assigned to the workers, increasing the probability that the final decision is correct. 
	\item However, the reward is not a strictly increasing function of $b$. This is because the increasing cost resulting from more questions undermines the gain in decision accuracy $R$. When the budget $b$ is sufficiently large, as observed from the two cases, $b = 13$ and $b = 17$, in \cref{fig:Budget}, the reward stays almost constant, since both POMDP and URSQS tend to stop and make a decision at a time before $b$ steps.
	\item We observe the reward for DCFECC could even decrease with increasing $b$. This is because even decision accuracy $R$ increases with increasing number of questions, it's not optimal to fully use the budgets $b$ of the crowdsourcer, i.e. to always declare at time $b$.
\end{itemize}

\subsection{Varying \texorpdfstring{$\gamma$}{gamma}}

\begin{figure}[!htb]
 \includegraphics[width=1\linewidth]{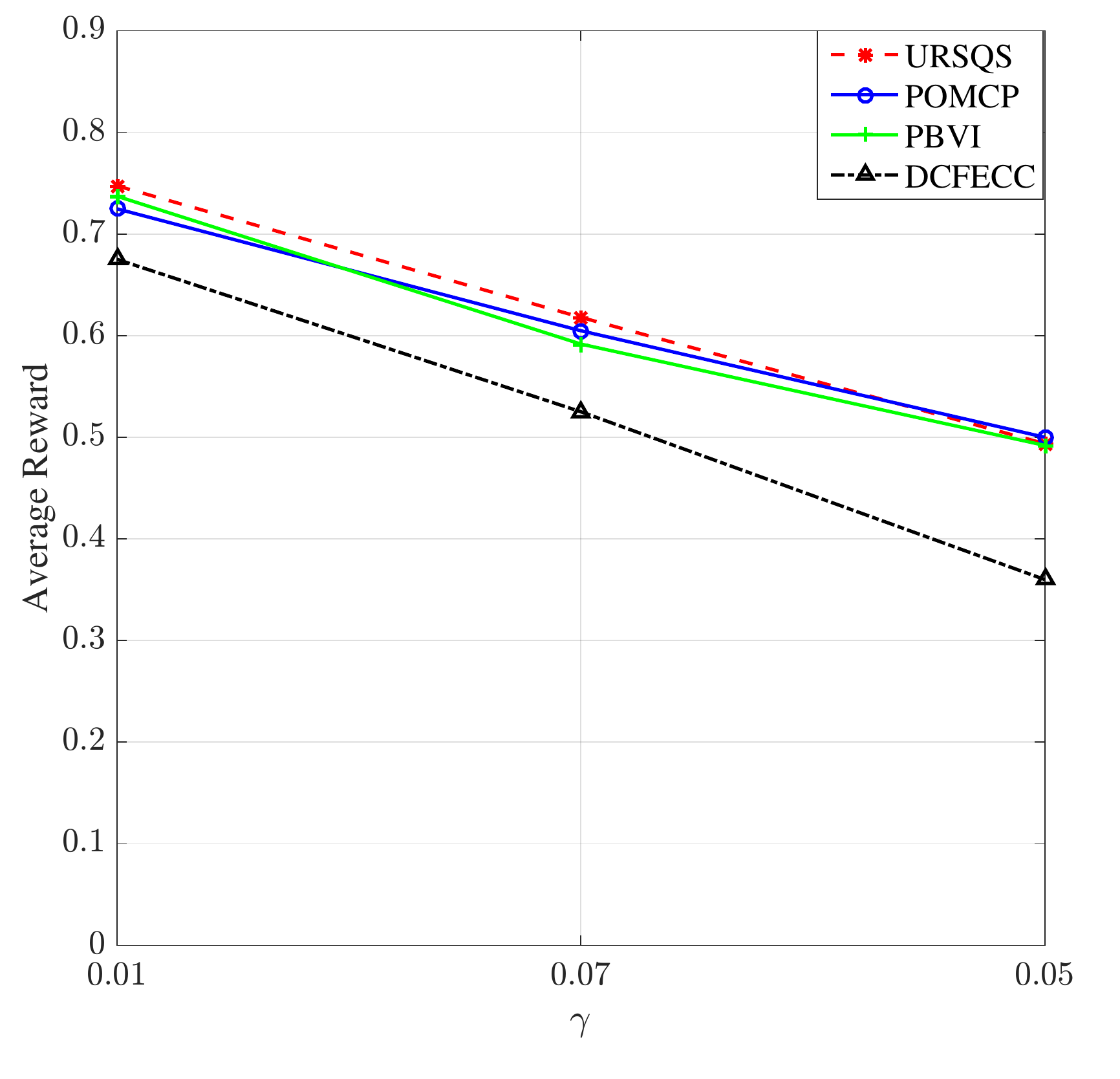}
 \vspace*{-8mm}
 \caption{Performance comparison between strategies with $M = 32$, $r=0.75$, $b = 9$ and different $\gamma$.}
 \label{fig:Cost}
\end{figure}

We next fix $M = 32$, $r = 0.75$, $b = 9$ and vary the cost $\gamma$ of posing a question. The results are shown in  \cref{fig:Cost}. We can see that the reward decreases with increasing $\gamma$. This is expected, since the larger cost will render the average reward lower.

\subsection{Varying \texorpdfstring{$r$}{r}}
\begin{figure}[!htb]
 \includegraphics[width=1\linewidth]{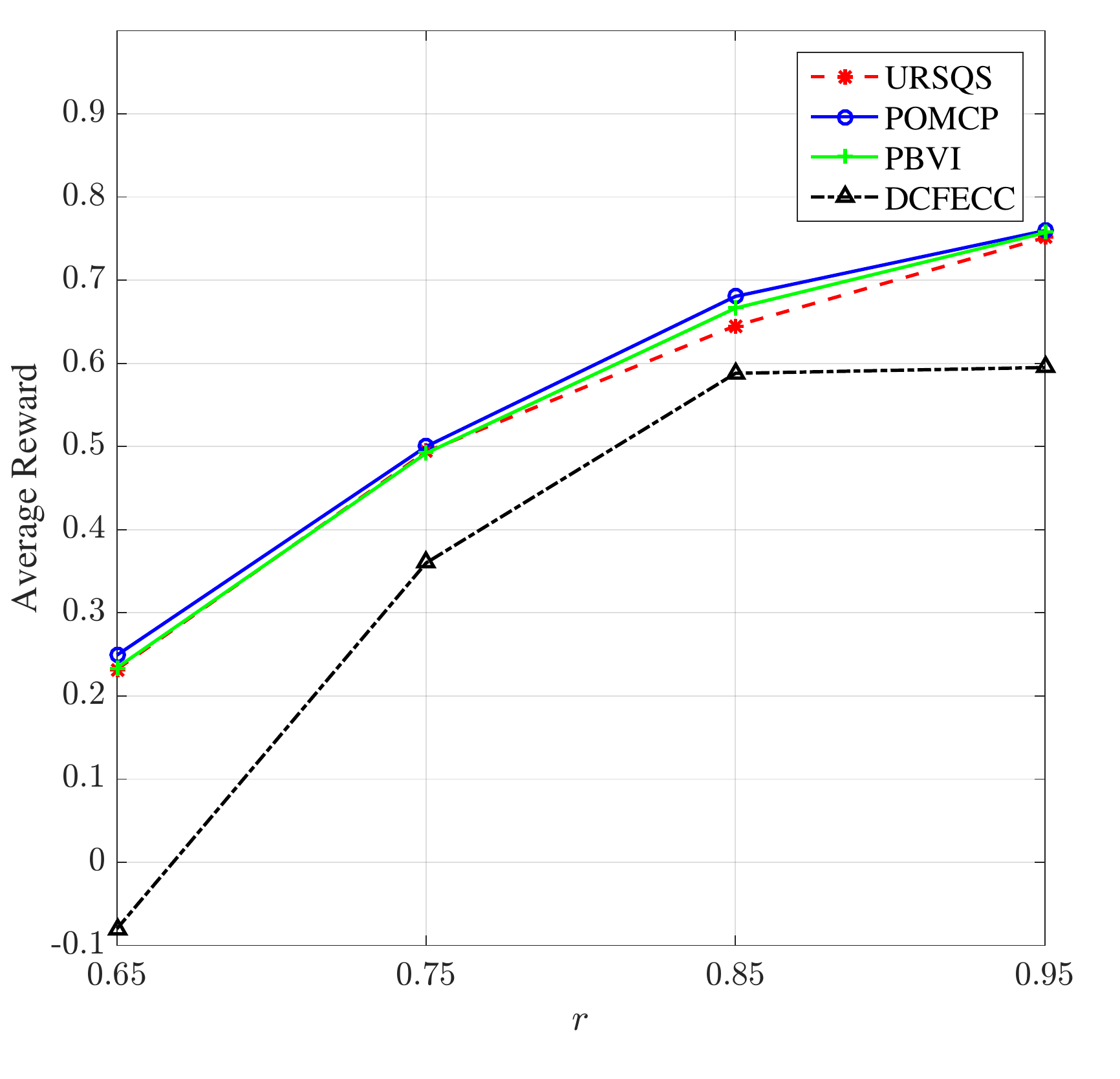}
 \vspace*{-8mm}
 \caption{Performance comparison between strategies with $M = 32$,  $b = 9$, $\gamma = 0.05$ and different $r$.}
 \label{fig:Reliability}
\end{figure}

We vary the parameter $r$ in $\mu_q$ with fixed $M = 32$, $b = 9$, and $\gamma = 0.05$. A larger $r$ implies a larger $\mu_q$, i.e., the workers are more reliable. From \cref{fig:Reliability}, we observe that the reward increases with increasing $r$, as the workers become more reliable. 

\subsection{Optimality of URSQS}

Finally, to verify the optimality of URSQS, we evaluate URSQS against the following benchmarks:
\begin{itemize}
	\item $L_{\text{URSQS}}(q) = \max_{e} L(q,e)$, for $q\in[2,M]$, and where $L(q,e)$ is the objective function of URSQS in \cref{ULObj} and the maximization is over all $e$ such that $\hat{B}(q,e) \leq b-1$.
	\item $U^*(q)$, for each $q\in[2,M]$, is the maximum reward achievable by our proposed Ulam-R\'enyi strategy in \cref{algor:heuris} over all $e$ such that $\hat{B}(q,e) \leq b-1$. This is found through an exhaustive search in each simulation trial, and serves as a benchmark to measure how well our URSQS formulation in \cref{ULObj} performs.
\end{itemize}

\begin{figure}[!htb]
  \includegraphics[width=1\linewidth]{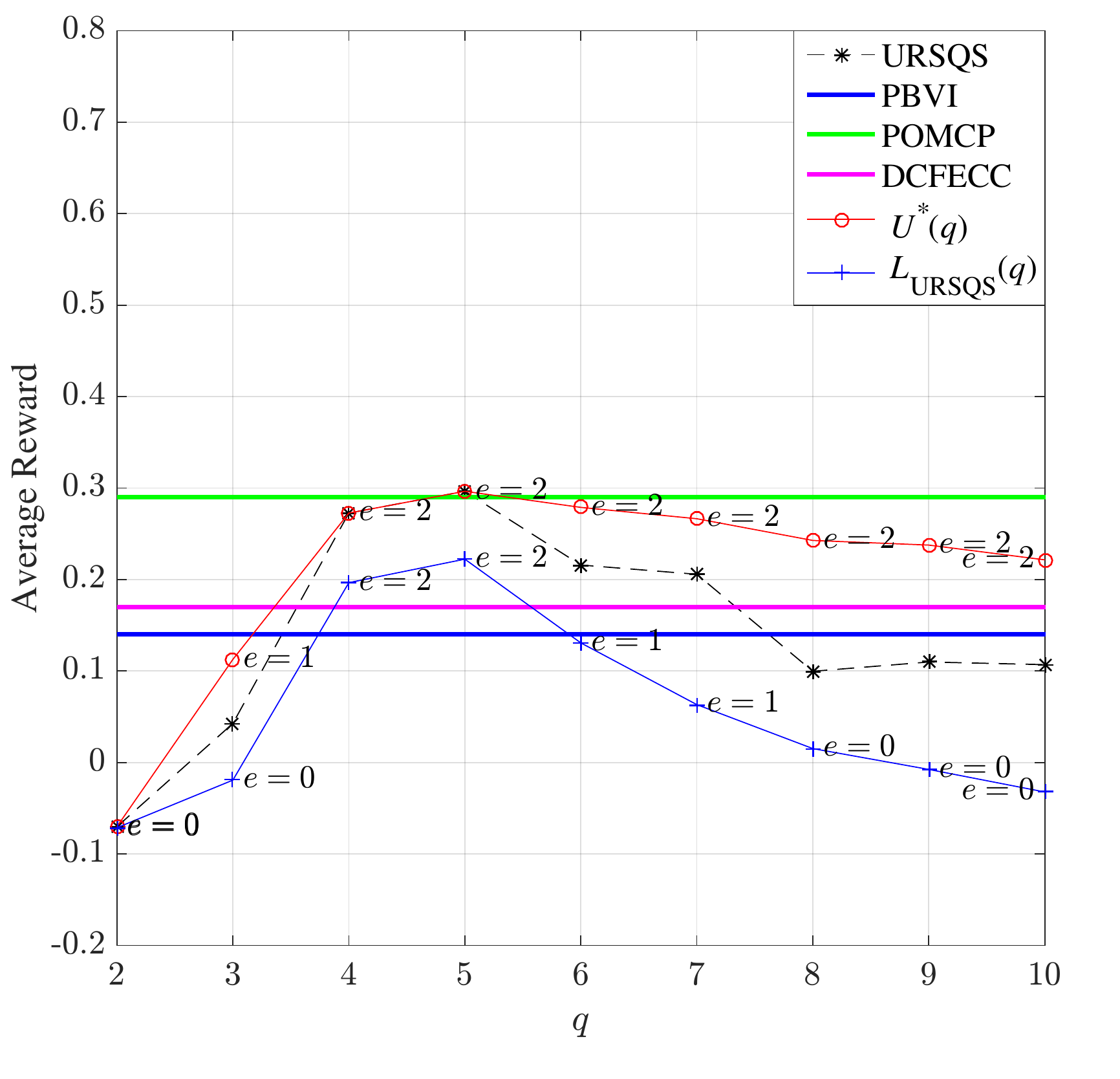}
	\vspace*{-8mm}
 \caption{Performance comparison between different strategies with $M = 128$, $r=0.75$, $b = 9$ and $\gamma=0.05$.}
  \label{fig:M128}
\end{figure}

\begin{figure}[!htb]
 \includegraphics[width=1\linewidth]{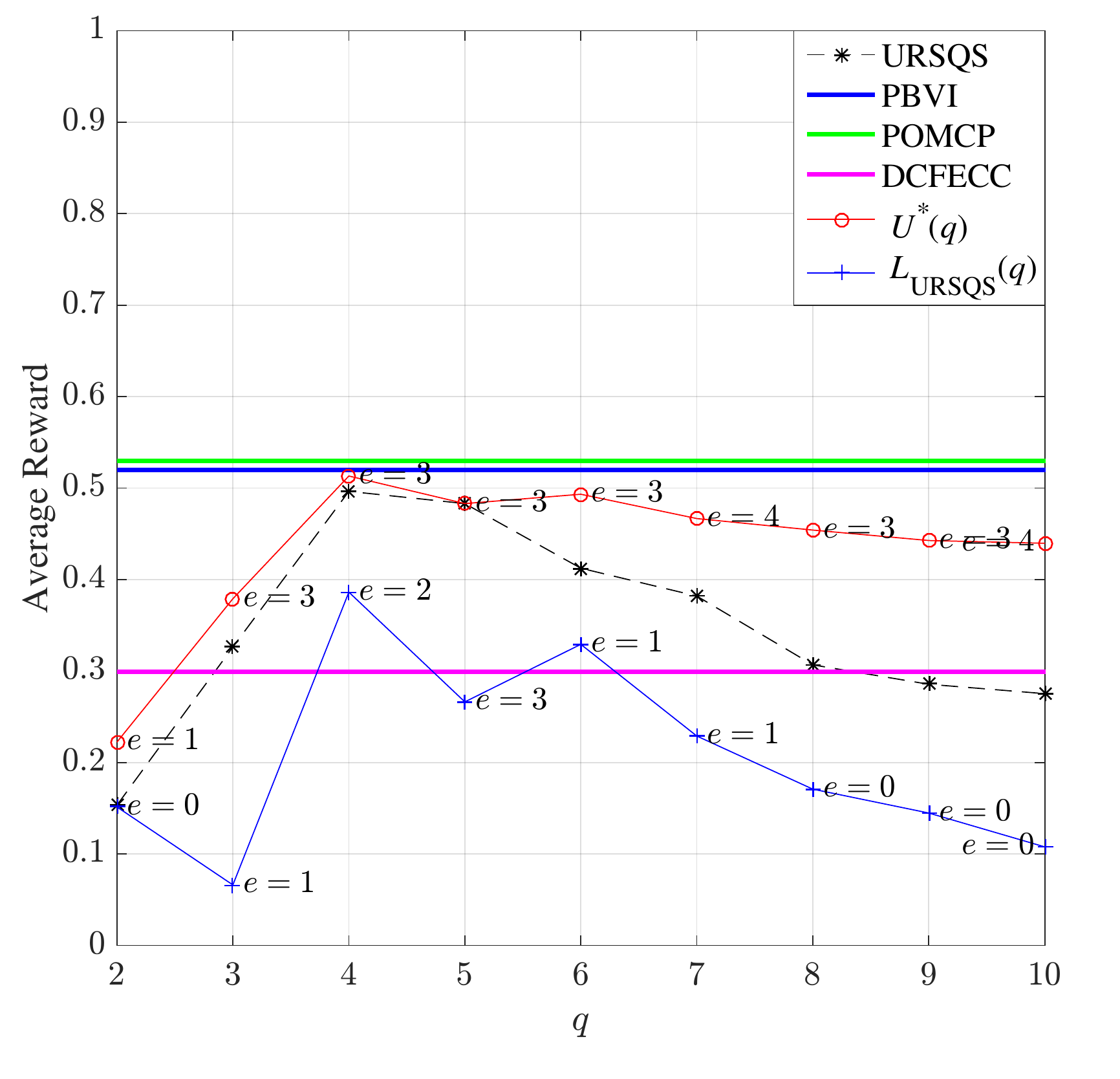}
 \vspace*{-8mm}
 \caption{Performance comparison between different strategies with $M = 32$, $r=0.75$, $b = 13$ and $\gamma=0.05$.}
 \label{fig:M32b13}
\end{figure}

We observe that in most of our experiments, $L_{\text{URSQS}}(q)$ achieves its maximum at the same $(q,e)$ as the maximum of $U^*(q)$, similar to that shown in \cref{fig:M128}. However, in some cases like \cref{fig:M32b13}, the maximum reward for URSQS is different from the optimal reward given by $U^*(q)$, but not significantly so. The average difference between the rewards obtained by URSQS and $U^*(q)$ is 0.017.

\section{Conclusion} \label{sec:conclusion}

We have developed sequential question design strategies for crowdsourcing in which workers' responses are used to determine the hidden state of a multi-class labeling problem. Our strategies allow the crowdsourcer to sequentially determine the question to pose to the workers, based on their previous responses. We formulated the sequential design as a POMDP, which maximizes a weighted function of the probability of inferring the correct hidden state after fusing the workers' responses, and the cost of posing additional questions, subject to a budget constraint for the crowdsourcer. To overcome the intractability of the POMDP, we have proposed design approaches based on the Ulam-R\'enyi game. We have demonstrated through simulations that when the number of hidden states is relatively large, our approaches achieve good performance, with less computational complexity when compared with traditional POMDP solvers.

In this paper, we have assumed that workers' responses are independent across questions and amongst the group of workers. A future direction is to consider the case where workers' responses may not be independent, and to study the joint optimization of what questions to pose to each worker, and how to fuse their responses together. Workers' previous responses to gold questions \cite{oleson2011programmatic} can also be used to infer their reliability, and to optimize the selection of workers for particular questions. It would be of interest to design a dynamic selection strategy that poses different questions to different workers optimally at each step.

\bibliographystyle{IEEEtran}
\bibliography{IEEEabrv,refs}
\begin{IEEEbiography}{Qiyu Kang}(S'17) received the B.S. degree in Electronic Information Science and Technology from University of Science and Technology of China in 2015. He is currently a Ph.D. candidate at the School of Electrical and Electronic Engineering, Nanyang Technological University. His research interests include distributed signal processing, collaborative computing, and social network.
\end{IEEEbiography}

\begin{IEEEbiography}{Wee Peng Tay}(S'06 M'08 SM'14) received the B.S. degree in Electrical Engineering and Mathematics, and the M.S. degree in Electrical Engineering from Stanford University, Stanford, CA, USA, in 2002. He received the Ph.D. degree in Electrical Engineering and Computer Science from the Massachusetts Institute of Technology, Cambridge, MA, USA, in 2008. He is currently an Associate Professor in the School of Electrical and Electronic Engineering at Nanyang Technological University, Singapore. His research interests include distributed inference and signal processing, sensor networks, social networks, information theory, and applied probability.

Dr. Tay received the Singapore Technologies Scholarship in 1998, the Stanford University President's Award in 1999, the Frederick Emmons Terman Engineering Scholastic Award in 2002, and the Tan Chin Tuan Exchange Fellowship in 2015. He is a coauthor of the best student paper award at the Asilomar conference on Signals, Systems, and Computers in 2012, and the IEEE Signal Processing Society Young Author Best Paper Award in 2016. He is currently an Associate Editor for the IEEE Transactions on Signal Processing, an Editor for the IEEE Transactions on Wireless Communications, serves on the MLSP TC of the IEEE Signal Processing Society, and is the chair of DSNIG in IEEE MMTC. He has also served as a technical program committee member for various international conferences.
\end{IEEEbiography}
\end{document}